\documentstyle[12pt]{article}

\def\be{\begin{eqnarray}}
\def\ee{\end{eqnarray}}
\def\nn{\nonumber}

\begin{document}

\hfill{ITEP-PH-1/97}

\hfill{TAC-1997-001}

\hfill{ March 5, 1997} \\

\bigskip

\centerline{\Large{DO MUONS OSCILLATE~?}}

\bigskip

\centerline{\it A.D.Dolgov $^{1,2)}$,
A.Yu.Morozov $^{1)}$, L.B.Okun $^{1)}$, M.G.Schepkin $^{1)}$}

\vspace{8mm}

\centerline{$^{1)}~$ ITEP, 117218, Moscow, Russia}

 \centerline{ $^{2)}~$Teoretisk Astrofysik Center}
\centerline { Juliane Maries Vej 30, DK-2100, Copenhagen, Denmark }

\bigskip

\centerline{Abstract}

\bigskip

We develop a theory of the EPR-like effects due to neutrino
oscillations in the $\pi \rightarrow \mu\nu$ decays.  Its
experimental implications are space-time correlations
of neutrino and muon when they are {\it both} detected, while the
pion decay point is not fixed.  However, the more radical
possibility of $\mu$--oscillations in experiments where
{\it only} muons are detected (as it has been suggested in
hep-ph/9509261), is ruled out. We start by discussing decays of
monochromatic pions, and point out a few ``paradoxes''.
Then we consider pion wave packets, solve the ``paradoxes'', and show
that the formulas for $\mu \nu$ correlations can be transformed into
the usual expressions, describing neutrino oscillations, as soon as
the pion decay point is fixed.

\bigskip

\bigskip

\section{Introduction}

It was suggested recently \cite{SSW,SWS} that
the much searched for hypothetical neutrino flavour oscillations can
cause space-time oscillations of the observation rate of the beam of
muons from the $\pi$-meson decay.  As emphasized in \cite{SSW}, the
same claim applies to other charged leptons, in particular to
$\tau$-leptons in the decay $W\rightarrow \tau \nu_\tau$.  If true,
this phenomenon could provide a powerful experimental method for the
(indirect) observation of neutrino oscillations.

We think that argumentation of \cite{SSW} is unfortunately erroneous,
and there are {\it no} oscillations of charged leptons if neutrinos
are not observed.  However, in coincidence experiments, in which both
the charged lepton and neutrino, born in the same decay, are
detected, specific EPR--like oscillating correlations \cite{EPR}
can show up (see Section 3.2).  Once neutrino
oscillations \cite{BMP1}--\cite{IYK} {\it per se} are discovered,
this could be of interest for the next generation of experiments.

If the neutrino mass matrix is not flavour diagonal then the
``current'' neutrinos $\nu_a$ ($a=e,\mu,\tau$) are non-trivial
mixtures of the mass eigenstates, $\nu_n$ ($n = 1,2,3$) with
masses $m_1 > m_2 >  m_3$:
\be
\nu_a = \sum_{n} U_{an} \nu_n,
\label{nuca}
\ee
i.e. the relevant part of the Standard Model Lagrangian is
\be
{\cal L}_\nu =
g \sum_a \bar l_a\hat W \nu_a + H.c =
g \sum_{a,n} U_{an}\bar l_a\hat W \nu_n + H.c.
\label{Lag}
\ee

In what follows we consider a ``toy model'' with only
two charged leptons, muon and electron, and two neutrinos,
$\nu_1$ and $\nu_2$:
\be
L_\nu = g \left(\bar\mu\hat W (\nu_1\cos\theta + \nu_2\sin\theta) +
\bar e \hat W (-\nu_1\sin\theta + \nu_2\cos\theta)\right)
+ H.c.
\label{Lagr}
\ee
Also, we neglect the width of the emerging muon (as the oscillations
discussed in ref.\cite{SSW} survive in the limit of stable $\mu$;
in any case all the widths are easy to restore in the formulas).

We consider the decay $\pi \rightarrow \mu\nu$,
and analyse three types of experiments:

$A$) when both neutrino and muon are detected; this is an experiment
with two detectors in coincidence;

$B$) when only muons are detected;

$C$) when only neutrinos are detected.

Evidently, $B$ and $C$ are single-detector experiments, and
$B$ is considerably easier than $C$ and $A$.

Our analysis will demonstrate that the probability to detect
both $\mu$ {\it and} $\nu$ (case $A$) oscillates as a function of the
distance $d$ between the muon and neutrino detection points and of
the time interval $\Delta t$ between ``clicks'' of the two detectors.
Moreover, oscillation length and oscillation frequency are different
from the standard \cite{BP} values $L = 2E_\nu / (m_1^2-m_2^2)$ and
$L^{-1}$, respectively.

These oscillations disappear completely in the case
$B$, just because of the orthogonality of different neutrino
mass eigenstates in the flavour space. However, they can
show up in the case $C$. There are two ways in which the
neutrino oscillations might manifest themselves: the first may be
called {\it global}, the second -- {\it local}.

To see the {\it global} effect one does not need to observe the
oscillating term: it is sufficient to observe appearance of
$\nu_e$, and/or disappearance of $\nu_\mu$. This could be done
without accurate measurements of the time when the decaying pion
was produced in the target, or the position of its decay point.
From the theoretical point of view the {\it global} effects
of appearance and disappearance can be described within the
approximation of plane waves (monochromatic pion).

To see the {\it local} effect, to observe time- and/or
space-oscillations, one needs an adequate resolution.
From the theoretical point of view this also requires
the pion to be not exactly monochromatic (which is automatically the
case in any realistic experiment). Otherwise the muons born together
with $\nu_1$ would be orthogonal to those born with $\nu_2$, and
the oscillation term would drop out from the total probability.
Although our conclusions coincide with the usual naive expectations,
we feel that it can be useful to present this analysis in a little
more detail, to avoid any further confusion in the
literature.\footnote{  Let us note that in the literature neutrino
oscillations are treated usually in an oversimplified way.
The superposition of two neutrino mass eigenstates is described by a
wave function, not by a density matrix. The two terms of the wave
function are usually assumed to be plane waves with the same momentum
${\bf p}_\nu$ and different energies,
$E_{\nu 1}= \sqrt{{\bf p}_\nu^2+m_1^2}$,
$E_{\nu 2}= \sqrt{{\bf p}_\nu^2+m_2^2}$
(see, e.g. books \cite{JNB,MP,CB,BV}), or with the same energy
$E_\nu$, and different momenta
$|{\bf p}_{\nu 1}|= \sqrt{E_\nu^2-m_1^2}$,
$|{\bf p}_{\nu 2}|= \sqrt{E_\nu^2-m_2^2}$
(see \cite{IYK,GrKl}, and especially 
\cite{Lipkin1,Lipkin2,Lipkin3}).
The 4-momentum conservation in the decay $\pi \to \mu \nu$
is usually ignored. Besides \cite{SSW,SWS}, there were only a few
papers \cite{RW,JKL,CJ}, in which possible
kinematic manifestations of the 4-momentum conservation  had
been discussed. The neutrino
wave packet had been considered only by few authors
\cite{AD,Kay,Nuss}.

After completion of this paper we have read
section 5.2 in ref.\cite{BV}, where momentum conservation and
wave packets were discussed in a spirit close to ours.}

In Part 2 we discuss the decay of a monochromatic pion described
by a plane wave, derive expression for correlations
$P^{(A)}_{\nu_a}(x_\mu, x_\nu)$ (Section 2.1), and analyse them
in two cases: ${\bf p}_\pi = 0$ (Section 2.2) and
${\bf p}_\pi \not= 0$ (Section 2.3). Then, in Section 2.4,
we show that if only one of the particles is detected, either
muon (case $B$), or neutrino (case $C$), the oscillating term
disappears. We argue that for muons the absence of
oscillations is natural, but for neutrinos it is an artifact
of the plane wave approximation for the pion. We also show that
the {\it global} effects of $\nu_e$--appearance and
$\nu_\mu$--disappearance are reproduced in the plane wave
approximation.

In Part 3 we use pion wave packet and derive expressions
for $\mu \nu$ system, as well as for separate beams of
muons and neutrinos (Section 3.1). These expressions are
extremely simple, as the oscillating terms depend only on the
time of flight of neutrino from its birth till its detection.
In Section 3.2 we rederive some of the results of Section 3.1
by using the technique of plane waves and of the so called
``classical {\it a posteriori} trajectories'' for muons and
neutrinos. Part 4 contains a brief summary and a few concluding
remarks.

\section{Decay of a monochromatic pion state}

\subsection{Experiment of the type $A$: The probability formula}

To begin with, let us assume that pion has a definite
4-momentum $p_\pi = (E_\pi,{\bf p}_\pi)$.
Then the 4-momenta $p_{\mu n}=(E_{\mu n},{\bf p}_{\mu n})$
and $p_{\nu n}=(E_{\nu n},{\bf p}_{\nu n})$ are determined by
the conservation law,
\be
p_\pi = p_{\mu n} + p_{\nu n}~,~~~~~n=1,2~,~~~~~
\label{cons}
\ee
and the {\it direction} of, say, ${\bf p}_{\nu n}$. In what follows
we assume that the momenta of all three particles are known: either
measured, or deduced from kinematics.  All particles are on the mass
shell:
\be
p_\pi^2 = m_\pi^2,~~~ p_{\mu 1}^2 = p_{\mu 2}^2 =
m_\mu^2, ~~~p_{\nu 1}^2 = m_1^2, ~~~ p_{\nu 2}^2 = m_2^2~.
\label{shell}
\ee

The wave function of the $\mu\nu$ system evolves in space-time as
\be
\psi_{p_\pi}(x_\mu,x_\nu|x_{i}) = ~~~~~~~~~~~~~~~~~~~
\nn \\ =
|\mu \rangle e^{-ip_{\mu 1}(x_\mu-x_{i})}
|\nu_1\rangle e^{-ip_{\nu_1}(x_\nu - x_{i})}\cos\theta +
\label{wf}    \\
+|\mu \rangle e^{-ip_{\mu 2}(x_\mu-x_{i})}
|\nu_2\rangle e^{-ip_{\nu_2}(x_\nu - x_{i})} \sin\theta=
\nn   \\    =
|\mu \rangle (|\nu_1\rangle e^{-i\varphi_1}\cos\theta +
|\nu_2\rangle e^{-i\varphi_2}\sin\theta)~,
\nn
\ee
where
\be
\varphi_1=p_{\nu 1}(x_\nu-x_i)+p_{\mu 1}(x_\mu-x_i)~,
\label{phis}     \\   \nn
\varphi_2=p_{\nu 2}(x_\nu-x_i)+p_{\mu 2}(x_\mu-x_i)~.
\ee
 $|\mu\rangle$, $|\nu_1\rangle$ and $|\nu_2\rangle$ are
``ket'' state vectors of the muon $\mu$ and neutrino mass eigenstates
$\nu_1$ and $\nu_2$ respectively; $p_{\nu 1}$ and $p_{\nu 2}$ are
4-momenta of $\nu_1$ and $\nu_2$, and $p_{\mu 1}$ and $p_{\mu 2}$
are those of muons emitted together with $\nu_1$ and $\nu_2$;
$x_\mu$ and $x_\nu$ are space-time coordinates of the muon and neutrino,
while $x_i$ is  the coordinate of the decay point ($i$ for initial).

The amplitude to detect a muon at a space-time point $x_\mu$
together with a neutrino of flavour $a$ ($a=e,~\mu$)
at the point $x_\nu$ is
\be
a_{\nu_a}(x_\mu,x_\nu) =  \langle \mu; \nu_a|
\psi_{p_\pi}(x_\mu,x_\nu)\rangle\ =
\beta_{1a} e^{-i\varphi_1} + \beta_{2a} e^{-i\varphi_2}~,
\label{anua}
\ee
and the corresponding probability is:
\be
P^{(A)}_{\nu_a} (x_\mu,x_\nu) =
|a_{\nu_a} (x_\mu, x_\nu)|^2 =
\label{PA}
   \\   \nn  =
\beta_{1a}^2 + \beta_{2a}^2 + 2\beta_{1a}\beta_{2a}
\cos (\varphi_1-\varphi_2)~,
\ee
where
\be
\varphi_1-\varphi_2=p_{\nu 1}(x_\nu-x_i) +p_{\mu 1}(x_\mu-x_i)
-p_{\nu 2}(x_\nu-x_i) -p_{\mu 2}(x_\mu-x_i)=
\label{phi12}
\\    \nn   =
(p_{\nu 1}-p_{\nu 2})(x_\nu-x_i)+
(p_{\mu 1}-p_{\mu 2})(x_\mu-x_i)=
\\    \nn    =
(p_{\nu 1}-p_{\nu 2})(x_\nu-x_i-x_\mu+x_i)
=(p_{\nu 1}-p_{\nu 2})(x_\nu-x_\mu)~.
\ee
We made use of the equality
$p_{\nu 1}-p_{\nu 2} = -~(p_{\mu 1}-p_{\mu 2})$,
which follows from the 4-momentum conservation (\ref{cons}).
Note that $\varphi_1-\varphi_2$ is explicitly Lorentz invariant.

It follows from (\ref{Lagr}), that
\be
\beta_{1\mu} = \cos^2\theta,  ~~~
\beta_{2\mu} = \sin^2\theta, ~~~
\beta_{2e} = -\beta_{1e} = \sin\theta\cos\theta~.
\label{beta}
\ee

Hence if the neutrino is detected as $\nu_\mu$,
\be
P^{(A)}_{\nu_{\mu}}(x_\mu,x_\nu) =
\cos^4\theta + \sin^4\theta + 2\sin^2\theta\cos^2\theta\cos
(\varphi_1-\varphi_2)~.
\label{PAmu}
\ee

If the neutrino is detected as $\nu_e$,
\be
P^{(A)}_{\nu_e}(x_\mu,x_\nu) =2\sin^2\theta\cos^2\theta -
2\sin^2\theta\cos^2\theta\cos(\varphi_1-\varphi_2)~.
\label{PAe}
\ee

The probability (\ref{PA}) oscillates in space and time with
the change of $x_\mu$ and/or $x_\nu$, presenting a kind
of EPR effect. Actually these oscillations depend only on
the differences ${\bf x}_\nu - {\bf x}_\mu$ and $t_\nu - t_\mu$,
and can be observed in an experiment, which detects both
muon and neutrino from the same decay.\footnote{
In fact, an experiment of this kind has been started at IHEP
(Protvino) by S.P.Denisov et al.  \cite{SPD,FS}.  They are looking for
(semi)leptonic decays of kaons: $K \to \mu \nu$, $K \to \mu\nu\pi$ and
$K \to e\nu\pi$. The energy of kaons is $E_K = 35 \pm 1$ GeV. The
experiment has two detectors.  The first detector measures muons and
electrons coming directly from the kaon decay; the second detector
measures muons and electrons produced by neutrinos. Thus, by
detecting a neutrino interaction they are able to determine the
momentum of electron or muon, which was born together with that
neutrino in kaon decay. In this way neutrinos are ``tagged''. By
analysing the kinematics they are able to reconstruct position of the
kaon decay point with an accuracy of 3 m.  The tagging provides
accurate information not only on the position of the origin of
neutrino, but also on its original flavour.  This allows, in
principle, a more accurate determination of the neutrino mixing angle
$\theta$.  }

The expression (\ref{PA}) was in fact derived but misinterpreted in
ref.\cite{SSW}.  An extra  $e^{-\Gamma_\mu t}$  in \cite{SSW}, which
takes into account the decay of muon, is not essential for our
analysis.

Before proceeding to the discussion of experiments $B$ and $C$,
let us look at eq.(\ref{PA}) a little closer by
substituting the expressions for $p_\mu$ and $p_\nu$
in two cases: ${\bf p}_\pi = 0$ and  ${\bf p}_\pi \not= 0$,
the former being a limiting case of the latter one. We start
with absolutely non-realistic case of
vanishing ${\bf p}_\pi$, because it had been discussed
in \cite{SSW}, because it has its own subtleties, and because
the pion rest-frame values of the muon and neutrino energies
enter some of the expressions for the more general case.

\subsection{Experiment of the type $A$:
pion with momentum strictly equal to zero}

In this situation all four particles ($\pi, \mu, \nu_1, \nu_2$)
are described by plane waves and are fully non-localized.
The wave function of the $\mu \nu$ system evolves in
space-time according to eqs.(\ref{wf}) and (\ref{phis}).

In the rest frame of the pion:
\be
E_{\nu n}^0=\frac{m_{\pi}^2-m_{\mu}^2+m_n^2}{2m_{\pi}}~,
\label{enu}
\ee
\be
E_{\mu n}^0=\frac{m_{\pi}^2+m_{\mu}^2-m_n^2}{2m_{\pi}}~,
\label{emu}
\ee
\be
{\bf p}_{\mu n}^0 = - {\bf p}_{\nu n}^0~.
\label{mom0}
\ee

After the neutrino and the muon are detected we may conclude that
pion has decayed somewhere on the line connecting two detection
points, but {\it a priori} we are in principle unable to indicate
the position of the pion decay point on this line.

As for the frequency of oscillations and the oscillation length,
they are determined from eqs.(\ref{enu})--(\ref{mom0}):
\be
\Delta E^0_\nu =
E^0_{\nu 1} - E^0_{\nu 2} = \frac{m_1^2 - m_2^2}{2m_{\pi}} =
 \frac{1}{L^0}~\frac{E^0_{\nu}}{m_{\pi}}~,
\label{e012}
\ee
\be
\Delta {\bf p}_\nu^0 =
{\bf p}_{\nu 1}^0 - {\bf p}_{\nu 2}^0~,~~
\nn
\ee
\be
{\bf v}_\nu^0 \Delta {\bf p}_\nu^0 \approx
- \frac{m_1^2 - m_2^2}{2E_\nu^0} \frac{E^0_{\mu}}{m_{\pi}}
=-\frac{1}{L^0}~\frac{E^0_{\mu}}{m_{\pi}}=
\Delta E^0_\nu-\frac{1}{L^0}~,
\label{p012}
\ee
where $L^0 = 2E^0_{\nu}/(m_1^2 - m_2^2)$ is the standard
oscillation length. Note, that $\Delta E_\nu^0$ and
$|\Delta {\bf p}_\nu^0|$ are not equal to
each other and to $1/L^0$. Thus, we have in this case in
eq.(\ref{PA})
\be \varphi_1-\varphi_2= \frac{E_\nu^0}{L^0m_\pi}\Delta t +
\frac{E_\mu^0}{L^0m_\pi}d~,
\label{dt0}
\ee
where $d$ is the distance between two detectors (detection points).
Taking $d$ to be fixed, and by measuring $\Delta t=t_\nu -t_\mu$, one
can deduce {\it a posteriori} the point on the line connecting two
detectors, where pion has decayed (see Section 3.2).

\subsection{Experiment of the type $A$: pion in flight with
strictly fixed (sharp) momentum\label{ultra}}

Let us consider now a beam of pions moving from left to right
along the line which connects the muon and neutrino
detectors.\footnote{The muon detector has to be to the left
of the neutrino detector because the latter must be shielded.}

If the velocity of a pion
${\bf v}_\pi={\bf p}_\pi/E_\pi$ ($v_\pi=|{\bf v}_\pi|$)
is low enough, $v_\pi < v_\mu^0$, where
 ${\bf v}^0_\mu={\bf p}_\mu^0/E_\mu^0$, both $\mu$ and $\nu$
 will be detected if pion decays ``inside'', between two
detectors. For $v_\pi > v_\mu^0$ both $\mu$ and $\nu$ will be
detected if pion decays ``outside'', to the left of the muon
detector. For $v_\pi = v_\mu^0$ pion must decay in the muon
detector.
Let us express $\varphi_1-\varphi_2$ through the time interval
$\Delta t = t_\nu - t_\mu$ and the distance between the
detectors ${\bf d}={\bf x}_\nu-{\bf x}_\mu$:
\be
\varphi_1-\varphi_2= \Delta E_\nu \Delta t
-{\bf d} \Delta {\bf p}_\nu =
(E_{\nu 1}-E_{\nu 2})\Delta t
-({\bf p}_{\nu 1}-{\bf p}_{\nu 2}){\bf d}~.
\label{12phi}
\ee

Lorentz transformations give
\be
E_{\nu n}=(E_{\nu n}^0+{\bf v}_\pi{\bf p}_{\nu n}^0)\gamma_\pi~,~~
{\bf p}_{\nu n}=({\bf p}_{\nu n}^0+{\bf v}_\pi E_{\nu n}^0)\gamma_\pi~,
\label{EPnu}
\\
E_{\mu n}=(E_{\mu n}^0+{\bf v}_\pi{\bf p}_{\mu n}^0)\gamma_\pi~,~~
{\bf p}_{\mu n}=({\bf p}_{\mu n}^0+{\bf v}_\pi E_{\mu n}^0)\gamma_\pi~,
\label{EPmu}
\ee
where $\gamma _\pi=1/\sqrt{1-v_\pi^2}=E_\pi/m_\pi$, and
thus\footnote{Let us note that the expressions for $\Delta E_\nu$
and $\Delta {\bf p}_\nu$, as given by eqs.(\ref{12phi}) and
(\ref{phitd}), differ from eqs.(5.8) and (5.9) for the same
observables in ref.\cite{BV}.}
\be
\varphi_1-\varphi_2=\frac{\gamma_\pi}{L^0m_\pi}
\left\{(E_\nu^0-v_\pi E_\mu^0)\Delta t+
(E_\mu^0-v_\pi E_\nu^0)d \right\}~.
\label{phitd}
\ee

If we take into account that
\be
L^0 = L \frac{E_\nu^0}{E_\nu}~,
\label{Lo}
\ee
we see that the oscillation frequency is proportional to
$E_\pi E_\nu /L$, while the oscillation length decreases as
$L/E_\pi E_\nu$. The explanation of this drastic dependence
which looks quite unexpected will be given in Section 3.2

\subsection{Experiments of the types $B$ and $C$ with
monochromatic pions \label{BC}}

Oscillations described by equations similar to
eqs.(\ref{PA})--(\ref{PAe}) and
(\ref{dt0}) were referred to in ref.\cite{SSW}
as oscillations of a muon beam. In order to clarify
the situation let us consider first the
detection of muons (without detecting neutrinos)
 in the case ${\bf v}_\pi = 0$.
The beam of muons is described not by a wave function, but by a
density matrix (see, e.g. \cite{LL}). The probability of the muon
detection is obtained by integrating eq.(\ref{PA}) over the neutrino
position ${\bf x}_{\nu}$ and summing over neutrino flavours. Each of
these operations results in the vanishing of the oscillating term in
eq.(\ref{PA}). Integration over ${\bf x}_{\nu}$ simply leaves no
dependence on ${\bf x}_{\mu}$. Summing over $\nu_a$ $(a = e, \mu)$
also eliminates oscillations since for the case of final $\nu_{\mu}$
and for the case of final $\nu_e$ the oscillating term enters with
the opposite signs, see eqs.(\ref{PAmu}), (\ref{PAe}).  Thus in the
case $B$ the probability of detecting a muon does not depend on ${\bf
x}_{\mu}$. Moreover, it does not depend on the muon momentum ${\bf
p}_\mu$. In the case ${\bf v}_\pi=0$ there is no beam of muons: they
come to the muon detector isotropically because the decaying pions
are fully de-localized due to uncertainty relation.

Note that the same refers to the case $C$: although the {\it global}
effects of $\nu_e$--appearance and $\nu_\mu$--disappearance are
present, the flux of neutrinos  after integration over
${\bf x}_{\mu}$ is isotropic, and the oscillating term is washed out.
The latter  conclusion may look to be in contradiction with the
generally accepted theory of neutrino oscillations. However the
problem is solved as soon as we take into account that in real
experiments we never have a decaying pion with sharp momentum. Its
momentum distribution has a non-vanishing coherent spread. As a
result pions are described by coherent packets of plane waves, and
are localized in space--time, \cite{GW}--\cite{SF}.

Let us now repeat this reasoning in the more formal way; namely,
come back to the probability formula (\ref{PA}) and
use it for formal description of experiments of the types
$B$ and $C$ in the general case of a plane wave with sharp
momentum ${\bf p}_\pi$. This is straightforward: one should just
sum over all the states of neutrino and muon, respectively.

We begin with the observation of muons (case $B$). The states of
neutrino are labeled by the ``flavour'' index $a=e, \mu$ and by the
``position'' ${\bf x}_\nu$. Thus the probability of the $\mu$
detection at the space-time point $x_\mu$, is equal to
\be
P^{(B)}_{\mu} (x_\mu) = \sum_a \int
P^{(A)}_{\nu_a}(x_\mu,x_\nu) d{\bf x}_\nu =
\label{PBdef}
\\       \nn   =
\int\left[P^{(A)}_{\nu_\mu}(x_\mu,x_\nu) +
P^{(A)}_{\nu_e}(x_\mu,x_\nu) \right]d{\bf x}_\nu~.
\ee
This can be alternatively formulated as a transition
from the wave function of $\mu\nu$ system to the density
 matrix \cite{LL} for $\mu$ given by
\be
\rho_\mu(x_\mu,x_\mu') = {\rm Tr}_\nu \int
\psi(x_\mu,x_\nu) \psi^{\ast}(x_\mu',x_\nu)d{\bf x}_\nu ~.
\label{rhoBdef}
\ee
Eq.(\ref{PBdef}) is identical to (\ref{rhoBdef}) for
$x_\mu = x_\mu'$. In accordance with the general principles of
quantum mechanics, the r.h.s. of (\ref{PBdef}) and (\ref{rhoBdef})
are automatically independent of $t_\nu$.

It remains to substitute (\ref{PA}) into (\ref{PBdef}).
Integration over ${\bf x}_\nu$ gives rise to
$\delta\left({\bf p}_{\nu_1} - {\bf p}_{\nu_2}\right) = 0$,
because $ {\bf p}_{\nu_1} \neq {\bf p}_{\nu_2}$. This is
already enough to eliminate the oscillation term.
However, it vanishes in experiment of the type $B$ for a more
fundamental reason: because of the summation over neutrino species.
Indeed, it is easy to see that the oscillation term cancels in
the sum of (\ref{PAmu}) and (\ref{PAe}) even {\it before}
integration over ${\bf x}_\nu$. Thus we get
\be
P^{(B)}_\mu(x_\mu) = 1~.
\label{PBCONST}
\ee

An attentive reader can get suspicious at this point. We identified
{\it two} reasons for elimination of oscillations in the case $B$:
orthogonality of neutrino the states in the flavour and momentum
spaces.  If we now turn to the observation of neutrinos  (case $C$),
the first reason is absent (a muon is always the same particle),
but the second reason is still present:
${\bf p}_{\mu 1} \neq {\bf p}_{\mu 2}$  -- and
this is enough to eliminate the neutrino oscillating term in the
case $C$, contrary to any reasonable expectations.

Still, we insist that in the ``sharp''
case under consideration this is really
true: the probability to observe neutrino $\nu_a$ at point $x_\nu$
\be
P^{(C)}_{\nu_a}(x_\nu) =  \int
P^{(A)}_{\nu_a}(x_\mu,x_\nu) d{\bf x}_\mu
\label{PCdef}
\ee
does {\it not} contain oscillating term if one substitutes
(\ref{PA}) into (\ref{PCdef}). If one uses eq.(\ref{PAmu}) for
$P^{(A)}_{\nu_\mu}(x_\mu,x_\nu)$ and eq.(\ref{PAe}) for
$P^{(A)}_{\nu_e}(x_\mu,x_\nu)$, one gets:
\be
P^{(C)}_{\nu_\mu}(x_\nu)= \beta_{1 \mu}^2+ \beta_{2 \mu}^2
= \cos^4\theta + \sin^4\theta < 1~,
\label{Cxxx}
\ee
\be
P^{(C)}_{\nu_e}(x_\nu)= \beta_{1e}^2+ \beta_{2e}^2 =
2\sin^2\theta \cos^2 \theta >  0~.
\label{Cyyy}
\ee

Thus, {\it global} manifestations of the oscillations,
the appearance of $\nu_e$ and disappearance of $\nu_\mu$,
are evident, but the oscillating terms themselves are absent.

There is a simple reason for this: our  assumption that the
decaying pion has a definite momentum, which is never true in
experiments. It is more or less obvious that allowing a minor
coherent dispersion in the momentum distribution of pion, one will
always find solutions to the equation
\be
{\bf p}_{\mu 1}({\bf p}_\pi) = {\bf
p}_{\mu 2}({\bar {\bf p}}_\pi)~.
\label{ppbarmu}
\ee
As we will see (footnote 8),
eq.(\ref{ppbarmu}) has a solution at $|\delta {\bf p}_\pi|=
|{\bf p}_\pi -{\bar {\bf p}}_\pi| \approx 1/(1-v_\pi)L$.

The momentum dispersion is necessarily present in any realistic
experiment: pions are usually localized in a region much smaller than
$1/|\delta {\bf p}_\pi|$, and then, by the uncertainty relation,
their momentum dispersion\footnote{To
avoid confusion, let us emphasize that here we speak about
the coherent dispersion of a pion produced in a given act of
collision with accompanying particles being in a given state.
This should not be mixed up with non-coherent momentum
dispersion of pions in the same beam. Non-coherent means
that the pion is produced with the same accompanying particles,
but being in a different state, or with other accompanying
particles, or produced in a different collision act.}
 should be much larger than $|\delta {\bf p}_\pi|$.

Thus we need to return to the very beginning and repeat our
analysis, allowing some small (as compared to the masses and
energies in the problem),  but non-vanishing momentum dispersion in
the wave function of the original pion. This is a simple calculation,
but still it deserves being done:  for example, one should check that
such dispersion does not wash out the oscillations in
experiments of the types $A$ and $C$. We shall also use this new
calculation to represent eq.(\ref{PA}) in a somewhat
different form; this can help one to better understand the
``paradoxical'' results of Section \ref{ultra}.

\section{Decay of a pion with small momentum dispersion}

\subsection{Experiments $A$, $B$, and $C$}

Let us now assume that pion has been created at a
space-time point $x_\pi$ with some momentum distribution
$\phi({\bf p}_\pi)$. This means that at the
decay point $x_i$ the pion wave function is
\be
\psi_\pi(x_i-x_{\pi}) =
\int  \phi({\bf p}_\pi)  e^{-ip_\pi(x_i-x_\pi)} d{\bf p}_\pi,
\ \ \ p_\pi^2 = m_\pi^2,
\label{psipack}
\ee
and that of the emerging $\mu\nu$ system,
\be
\Psi(x_\mu,x_\nu) =
\int  \phi({\bf p}_\pi)  e^{-ip_\pi(x_i-x_\pi)}
\psi_{p_\pi}(x_\mu,x_\nu|x_i) d{\bf p}_\pi,
\label{WF}
\ee
with $\psi_{p_\pi}(x_\mu,x_\nu|x_i)$ given by eq.(\ref{wf}).

Actually $\Psi(x_\mu,x_\nu)$ does not depend on $x_i$,
because we imposed the conservation law (\ref{cons}) in all
our formulas.\footnote{The usual logic in quantum field theory
is reversed: one includes integration $\int d^4x_i$
in the definition (\ref{WF}) of $\Psi(x_\mu,x_\nu)$, and
this integration {\it leads} to the conservation law (\ref{cons}).
Our presentation is organized to minimize the length of the
reasoning. We do not need to be too careful about such
details; note that we ignore all the normalization factors,
irrelevant for the main topic of the paper. }
We shall, however, proceed a little differently and keep
$x_i$ for a while. Moreover, we introduce the condensed
notations: $x_{i\pi} = x_i - x_{\pi}$,
$x_{\mu i} = x_\mu - x_i$, and $x_{\nu i} = x_\nu - x_i$.

By using eq.(\ref{WF}) we define the amplitude
${\cal A}_{\nu_a}(x_\mu,x_\nu)$ (compare with eq.(\ref{anua})
for $a_{\nu_a}(x_\mu,x_\nu)$):
\be
{\cal A}_{\nu_a}(x_\mu,x_\nu) =
\int d{\bf p}_\pi  \phi({\bf p}_\pi) \sum_n \beta_{na}
\exp \left({-ip_\pi x_{i\pi} - ip_{\mu n} x_{\mu i} -
ip_{\nu n} x_{\nu i} }\right)~,
\ee
and get for the probability:
\be
{\cal P}^{(A)}_{\nu_a}(x_\mu,x_\nu) =
|{\cal A}_{\nu_a}(x_\mu,x_\nu)|^2 =
\label {PPA0}
\\  \nn  =
\int \int d{\bf p}_\pi d\bar{\bf p}_\pi
 \phi({\bf p}_\pi) {\phi(\bar{\bf p}_\pi)}
\sum_{n,\bar n}  \beta_{na} \beta_{\bar n a}
e^{-i(\varphi_n - {\bar\varphi}_{\bar n})}~,
\ee
where
\be
\varphi_n - {\bar\varphi}_{\bar n}=
(p_\pi - \bar p_\pi) x_{i\pi} +
(p_{\mu n} - \bar p_{\mu \bar n}) x_{\mu i} +
(p_{\nu n} - \bar p_{\nu \bar n}) x_{\nu i}~,
\label{ppbar}
\ee
and $n(\bar n) = 1,~2$.

Let us now make use of the assumption that dispersion of
the distribution $\phi({\bf p}_\pi)$ is small as compared
to the typical energies in the problem. This allows one to put
\be
\delta E_\pi \equiv E_\pi - \bar E_\pi =
\frac{1}{E_\pi}{\bf p}_\pi({\bf p}_\pi - \bar{\bf p}_\pi) =
{\bf v}_\pi ({\bf p}_\pi - \bar{\bf p}_\pi)
\equiv {\bf v}_\pi \delta {\bf p}_\pi~, \nn \\
\delta E_\mu \equiv  E_{\mu n} - \bar E_{\mu \bar n} =
\frac{1}{E_\mu}{\bf p}_\mu({\bf p}_{\mu n} - \bar{\bf p}_{\mu \bar n}) =
{\bf v}_\mu ({\bf p}_{\mu n} - \bar{\bf p}_{\mu \bar n})
\equiv {\bf v}_\mu \delta {\bf p}_\mu~,
\label{pimunu}
 \\ \nn
\delta E_\nu \equiv E_{\nu n} - \bar E_{\nu \bar n} =
\frac{1}{E_\nu}\left(
{\bf p}_\nu({\bf p}_{\nu n} - \bar{\bf p}_{\nu \bar n})
+\frac{m_n^2 - m_{\bar n}^2}{2}\right) =
\\     \nn =
{\bf v}_\nu ({\bf p}_{\nu n} - \bar{\bf p}_{\nu \bar n}) +
\frac{m_n^2 - m_{\bar n}^2}{2E_\nu} \equiv
{\bf v}_\nu \delta {\bf p}_\nu +\frac{m_n^2-m_{\bar n}^2}{2E_\nu} ~,
\ee
where ${\bf v}$'s are velocities of the particles,
${\bf v}={\bf p}/E$.
(Formulas (\ref{pimunu}) are derived by subtraction of
the mass-shell  equalities $E^2 = {\bf p}^2 + m^2$ and
${\bar E}^2 = {\bar {\bf p}}^2 + m^2$).  Note also that
from energy conservation one gets
\be
\delta E_\pi = \delta E_\mu + \delta E_\nu~.
\label{EEE}
\ee

Substituting this into (\ref{ppbar}) we get:
\be
\varphi_n - {\bar\varphi}_{\bar n}=
- ({\bf p}_\pi - \bar{\bf p}_\pi)({\bf x}_{i\pi}
 - {\bf v}_\pi t_{i\pi}) -
 \nn   \\          -
({\bf p}_{\mu n} - \bar{\bf p}_{\mu \bar n})
({\bf x}_{\mu i} - {\bf v}_\mu t_{\mu i}) -
({\bf p}_{\nu_n} - \bar{\bf p}_{\nu_{ \bar n}})
({\bf x}_{\nu i} - {\bf v}_\nu t_{\nu i}) +
\label{PPA}
 \\    \nn +~
\frac{m_n^2 - m_{\bar n}^2}{2E_\nu} t_{\nu i}~.
\ee

From eq.(\ref{PPA}) we may derive formulas for the cases $A,B,C$
with the pion being described by a wave packet.

When considering the case $A$, let us note that the first three
terms in eq.(\ref{PPA}) vanish on the trajectories of the particles,
\be
{\bf x}_{i\pi} = {\bf v}_\pi t_{i\pi}, ~~~
{\bf x}_{\mu i} = {\bf v}_\mu t_{\mu i}, ~~~
{\bf x}_{\nu i} = {\bf v}_\nu t_{\nu i}~.
\label{traj}
\ee
Hence\footnote{ Let us check again that the phase difference
given by eqs.(\ref{PPA})--(\ref{phixxx})
is Lorentz invariant. In the reference
frame, which moves with the velocity $\bf u$,
$t_{\nu i} \to \gamma_u (t_{\nu i}+{\bf ux}_{\nu i})=
(1+{\bf uv}_\nu)\gamma_u t_{\nu i}$ and
$E_\nu \to \gamma_u (E_\nu + {\bf up}_\nu)=
(1+{\bf uv}_\nu)\gamma_u E_\nu$, where
$\gamma_u = 1/\sqrt{1-{\bf u}^2}$, so that
$t_{\nu i}/L = (t_{\nu i}/E_\nu)(m_1^2-m_2^2)/2$
remains invariant.}
\be
\varphi_n - {\bar\varphi}_{\bar n}~=~ \frac{t_{\nu i}}{L}~
~~~~\mbox{\rm for}~~~ n=1,~~\bar n = 2~.
\label{phixxx}
\ee

Note that in the case $A$ the muon detector is used in order to
deduce the position in space-time of the pion decay point $i$.
Hence the value of $t_{\nu i}$ (in the case $A$ we may call it
$t_{\nu i}^A$) can be determined
without the knowledge of the moment of the pion production
$t_\pi$ ($x_\pi = (t_\pi,{\bf x}_\pi)$, see eq.(\ref{psipack})).
This will be done explicitly in eqs.(\ref{xxA})--(\ref{tnui2}).
Our new expression for the probability in the case $A$,
\be
{\cal P}^{(A)}_{\nu_a}(x_\mu,x_\nu) =
\sum_{n,\bar n}  \beta_{na} \beta_{\bar n a}
\exp \left (i \frac{m_n^2 -
m_{\bar n}^2}{2E_\nu} t_{\nu i} \right )=
\nn \\ =
\label{PAn}
\beta_{1a}^2 + \beta_{2a}^2 + 2\beta_{1a}\beta_{2a}
\cos\frac{t_{\nu i}^A}{L}
\ee
looks absolutely different from the old one (compare
(\ref{PPA}) and (\ref{PAn}) with (\ref{PA}), (\ref{phi12}),
(\ref{dt0}), (\ref{phitd})).  In the next Section we will
show that in fact they are equivalent, and
$\varphi_1-\varphi_2 = \varphi_1-{\bar \varphi}_2
=t_{\nu i}/L$.

The same momentum dispersion technique can be applied to the
description of realistic experiments of the types $B$ and $C$.
In the case $C$ (single neutrino detector) by integrating
(\ref{PPA0}) over ${\bf x}_\mu$, one gets delta-function
$\delta({\bf p}_{\mu n} - \bar{\bf p}_{\mu \bar n})$,
where ${\bf p}_{\mu n}$ and ${\bar {\bf p}}_{\mu \bar n}$
are expressed through ${\bf p}_\pi$ and ${\bar {\bf p}}_\pi$,
respectively. If $n \neq \bar n$, the argument of the
delta-function vanishes for non-vanishing
${\bf p}_\pi - \bar {\bf p}_\pi = \delta {\bf p}_\pi$,
which for high energy pions must be equal to
$2\gamma^2_\pi/L = \gamma_\pi/L^0$
(see also eq.(\ref{phitd})).\footnote{To find exact expression
for ${\bf p}_\pi - \bar {\bf p}_\pi= \delta {\bf p}_\pi$, let
us consider ${\bf p}_\mu$ as a function of ${\bf p}_\pi$ and
$m_\nu$, and calculate the difference
$\delta {\bf p}_\mu=\delta {\bf p}_\pi - \delta {\bf p}_\nu$.
By using eqs.(\ref{pimunu}) and (\ref{EEE}) it is easy
to find $\delta {\bf p}_\pi$, corresponding to
$\delta {\bf p}_\mu={\bf p}_{\mu n}-\bar{\bf p}_{\mu \bar n}=0$
(and hence $\delta E_\mu=0$). The condition
$\delta {\bf p}_\mu=0$ means
$\delta {\bf p}_\nu=\delta {\bf p}_\pi$ and
$\delta E_\nu= \delta E_\pi$, hence
$({\bf v}_\pi - {\bf v}_\nu)\delta {\bf p}_\pi=1/L$, where we
put $\delta m_\nu^2 = m_1^2-m_2^2$. For the collinear case,
${\bf v}_\nu~||~{\bf v}_\pi$, we get
$|\delta {\bf p}_\pi| = L^{-1}/(1-v_\pi)$ since
$v_\nu \approx 1$. For high energy pions $|\delta {\bf p}_\pi|$
is much larger than $L^{-1}$: $|\delta {\bf p}_\pi|
\approx 2\gamma^2_\pi/L = \gamma_\pi /L^0$.}
The corresponding probability is
\be
{\cal P}^{(C)}_{\nu_a}(x_\nu)  =
\beta_{1a}^2 + \beta_{2a}^2 + 2\xi\beta_{1a}\beta_{2a}
 \cos\frac{t_{\nu i}^C}{L},
\label{PCn}
\ee
where
\be
\xi = \frac{ \int d{\bf p}_\pi
\phi({\bf p}_\pi-\delta {\bf p}_\pi /2)
\phi({\bf p}_\pi + \delta {\bf p}_\pi /2) }
{\int d{\bf p}_\pi {\phi}^2({\bf p}_\pi) }~,
\label{xi}
\ee
and $t_{\nu i}^C$ can be expressed through $x_\nu$ and
$x_\pi$, the space-time point where pion has been created;
in the collinear case (${\bf p}_\nu || {\bf p}_\pi$):
\be
t_{\nu i}^C =
\frac{|{\bf x}_\nu-{\bf x}_\pi|-v_\pi(t_\nu-t_\pi)} {v_\nu- v_\pi}~.
\ee

For monochromatic pions $\xi = 0$, and
\be
{\cal P}^{(C)}_{\nu_a}(x_\nu|\xi = 0) =
\beta_{1a}^2 + \beta_{2a}^2~,
\label{PCxi}
\ee
while in realistic experiments $\xi = 1$, and
\be
{\cal P}^{(C)}_{\nu_a}(x_\nu|\xi = 1) =
\beta_{1a}^2 + \beta_{2a}^2 +
2\beta_{1a}\beta_{2a}\cos\frac{t_{\nu i}^C}{L}~.
\label{PCxi0}
\ee

The value of $t_{\nu i}^C$ in eq.(\ref{PCn}) can be fixed only if the
moment of the production of a pion in target, $t_\pi$, is known with
a very high accuracy. For that purpose the time-structure of the
proton beam should be of the order of few nanoseconds. Otherwise the
oscillating term in eq.(\ref{PCn}) would be washed out, and only
{\it global} manifestations of neutrino oscillations
($\nu_e$--appearance and $\nu_\mu$--disappearance) would remain;
compare with eqs.(\ref{Cxxx}) and (\ref{Cyyy}).
Another obvious way to fix $t_{\nu i}^C$ and thus to observe
the oscillating term is to ascertain the position of the pion
decay point $x_i$.

As for the case $B$, it has been already mentioned that the
oscillating term vanishes after summation over neutrino flavours (see
Section 2.4). This statement is obviously true in the situation when
pion is described by a packet of plane waves:  summation over
$a = e,\mu$ using eq.(\ref{PAn})  gives ${\cal P}^{(B)}(x_\mu)= 1$.

\subsection{On equivalence of the two representations for $P^{(A)}$
\label{dop}}

There are two main ingredients in the previous section, namely:
1) the momentum representation of the pion wave packet (eq.(\ref{WF})),
and 2) the classical trajectories for fast particles, eq.(\ref{traj}).
Actually we could use only the second one in order to get eq.({\ref{PAn}),
and to solve the ``paradoxes'' of sections 2.2 and 2.3.

In this section we will work with plane waves and at the same time
with classical trajectories for $\mu$ and $\nu$ (see e.g. \cite{GW}).
After $\mu$ and $\nu$ are detected at certain space-time ``points''
with given momenta, we have, so to say,
``{\it a posteriori} packets'' of these particles, for which
${\bf x}_{\mu i} = {\bf v}_\mu t_{\mu i}$,
${\bf x}_{\nu i} = {\bf v}_\nu t_{\nu i}$, where
${\bf v}_\mu ={\bf p}_\mu/E_\mu$, and
${\bf v}_\nu ={\bf p}_\nu/E_\nu$.
Note that the wave length of $\mu$ or $\nu$ does not exceed
$10^{-13}$ cm, while the characteristic distances in neutrino
experiments are larger than hundred meters.

For $\nu$ and $\mu$ on the mass-shell:
\be
\Delta E_\nu = E_{\nu 1}-E_{\nu 2}=
{\bf v}_\nu \Delta {\bf p}_\nu +\frac{m_1^2-m_2^2}{2E_\nu}~,
\label{DENU}
\ee
\be
\Delta E_\mu = E_{\mu 1}-E_{\mu 2}= {\bf v}_\mu \Delta {\bf p}_\mu ~,
\label{DEMU}
\ee
where
\be
\Delta {\bf p}_\nu = {\bf p}_{\nu 1}-{\bf p}_{\nu 2}~,~~~~~
\Delta {\bf p}_\mu = {\bf p}_{\mu 1}-{\bf p}_{\mu 2}~.
\ee
Note that if the pion momentum is sharp,
\be
\Delta E_\pi =0~, ~~~~~\Delta {\bf p}_\pi =0~,
\label{EpiPpi}
\ee
then
\be
\Delta E_\mu = - \Delta E_\nu ~,~~~~~
\Delta {\bf p}_\mu = - \Delta {\bf p}_\nu~,
\label{DDMUNU}
\ee
and hence
\be
{\bf v}_\nu \Delta {\bf p}_\nu  +
 {\bf v}_\mu \Delta {\bf p}_\mu =
 ({\bf v}_\nu - {\bf v}_\mu) \Delta {\bf p}_\nu =
\frac{m_1^2-m_2^2}{2E_\nu} = \frac{1}{L}~.
\ee

In eq.(\ref{phixxx}) we have shown that
$\varphi_1-{\bar \varphi}_2=t_{\nu i}/L$.
Consider now $\varphi_1-\varphi_2$ as given by eq.(\ref{phi12}),
and use eqs.(\ref{DENU}), (\ref{DEMU}), and (\ref{DDMUNU}):
\be
\varphi_1-\varphi_2=
(p_{\nu 1}-p_{\nu 2})(x_{\nu i}-x_{\mu i})=
\nn   \\     =
\Delta E_\nu(t_{\nu i}-t_{\mu i})-
\Delta {\bf p}_\nu({\bf x}_{\nu i}-{\bf x}_{\mu i})=
\label{FI12}\\     \nn =
t_{\nu i}(\Delta E_\nu - {\bf v}_\nu \Delta {\bf p}_\nu)+
t_{\mu i}(\Delta E_\mu- {\bf v}_\mu \Delta {\bf p}_\mu) =
\frac{t_{\nu i}}{L}~.
\ee

Thus $\varphi_1-{\bar \varphi}_2 = \varphi_1- \varphi_2$,
as was promised after eq.(\ref{PAn}).
Eq.(\ref{FI12}) directly brings us from eq.(\ref{PA}) to
eq.(\ref{PAn}). We would get the same result in the cases
of a pion with ${\bf p}_\pi=0$  (eq.(\ref{dt0})), and of a
relativistic pion (eqs.(\ref{12phi})-(\ref{phitd})).
At fixed distance between the detectors $d$ the measurement
of $\Delta t$~~(time difference between ``clicks'' of the
two detectors) allows one to find the space-time point
$x_i$ of the pion decay. Thus for a pion with $v_\pi > v_\mu^0$
decaying to the left of the muon detector we have in the
collinear case discussed in Section 2.3
\be
|{\bf x}_\mu - {\bf x}_i| =  v_\mu(t_\mu - t_i), ~~~
|{\bf x}_\nu - {\bf x}_i| = v_\nu(t_\nu - t_i)~,
\label{xxA}
\ee
hence
\be
t_i = \frac{-|{\bf x}_\nu - {\bf x}_\mu| -
v_\mu t_\mu + v_\nu t_\nu}{v_\nu - v_\nu}~,
\ee
and
\be
t_{\nu i}^A = t_{\nu i} = t_\nu - t_i =
\frac{v_\mu (t_\mu - t_\nu)}{v_\nu - v_\mu} +
\frac{|{\bf x}_\nu - {\bf x}_\mu|}{v_\nu - v_\mu}=
\frac{-v_\mu \Delta t +d}{v_\nu-v_\mu}~.
\label{tnui1}
\ee

Taking into account that $v_\nu \approx 1$, we see that
in the last equation denominators are extremely small for
ultrarelativistic muons. That means that the decay of the pion
discussed in Section 2.3 must take place at very large
distance  from the detectors, unless the numerator in
eq.(\ref{tnui1}) is also very small.

For a pion with $v_\pi < v_\mu^0$, decaying between the
two detectors
\be
t_{\nu i}^A=t_{\nu i} = \frac{v_\mu \Delta t +d}{v_\mu+v_\nu} ~.
\label{tnui2}
\ee

It is important to stress that in eq.(\ref{PPA}) or
(\ref{FI12}) the phase $\varphi_1-\varphi_2$ depends only
on $t_{\nu i}$ and does not depend on $t_{\mu i}$. Thus these
formulas simply describe the standard neutrino oscillations
from the point of creation of the neutrino till the point
of its detection. Accordingly they do not depend on the position
of the muon detector. Therefore EPR--like correlations between
muon and neutrino detection appear only when we express $t_{\nu i}$
in terms of $\Delta t = t_\nu - t_\mu$ and
${\bf d}={\bf x}_\nu-{\bf x}_\mu$. If we assume that a special
detector measures the decay point of the pion $x_i$, then the
situation becomes absolutely trivial.

\section{Concluding remarks}

Let us summarize the main results.

In consideration of decay $\pi \to \mu\nu$ and subsequent neutrino
oscillations we assumed that the 4-momentum is conserved and all
particles ($\pi$, $\mu$, $\nu$'s) are on the mass-shell.

We used two different approximations:

1. The pion is monochromatic (has definite, sharp momentum).
Then the same is true for $\mu$ and $\nu$'s (up to an ambiguity
in rotation of the $\mu\nu$ plane, which is inessential for our
purposes). Thus all particles are described by plane waves.
Non-trivial phenomena, associated with neutrino oscillations, are due
to a slight difference between the 4-momenta of neutrinos with
different masses.  As a corollary, the muon, created in decay of the
same pion together with different neutrinos, also possesses slightly
different 4-momenta.

2. The pion is localized in space and is described by a (coherent)
wave packet.  Then the $\mu\nu$ wave function is a linear
superposition of wave packets, resulting from decays of pion at
different times.  It is reduced to a product of packets for $\mu$ and
$\nu$, when {\it any} of the particles is detected at a definite
space-time point -- this gives rise to what we call ``classical {\it
a posteriori} trajectories''.

We analyzed the following three experimental situations.

$A$. Muon and neutrino detected in coincidence.
The probability of detecting $\mu$ at a space-time point
$x_\mu$ and $\nu_a$ ($a=e,\mu$) at a point $x_\nu$ is given by
eq.(\ref{PAn}), where $t_{\nu i}^A$ is defined by eq.(\ref{tnui1})
or (\ref{tnui2}).

$B$. Neutrino is ignored, while $\mu$ is detected at the space-time
point $x_\mu$. The corresponding probability does not depend on
$x_\mu$ at all, ${\cal P}^{(B)}(x_\mu) = 1$,
and no traces of neutrino oscillations are seen.

$C$. Muon is ignored, while $\nu_a$ is detected at the space-time
point $x_\nu$. This time the probability depends essentially on the
spread of the (unobserved) muon wave packet and thus on that of
original pion. For monochromatic pion or, more precisely,  for
$|\delta {\bf p}_\pi| \ll \gamma_\pi^2 / L$ in a collinear
case, the parameter $\xi$ in
eq.(\ref{PCn}) vanishes, and there is no oscillating $x_\nu$
dependence in the probability, see eq.(\ref{PCxi}).
The only {\it global} consequence of neutrino oscillation in this
case is that ${\cal P}_{\nu_e}^{(C)} > 0$, while
${\cal P}_{\nu_\mu}^{(C)} < 1$.  For large enough\footnote{Note that
for ultrarelativistic pions, and in the collinear approximation,
$\delta {\bf p}_\pi$ has to be many orders of
magnitude larger than the naive estimate,
$|\delta {\bf p}_\pi| \gg  1/L$ !} dispersion,
$|\delta {\bf p}_\pi| \gg \gamma_\pi^2/L$, we have $\xi = 1$,
and the standard expression
${\cal P}^{(C)}_{\nu_a}(x_\nu)$ given by eq.(\ref{PCxi0}).
If $t_\pi$ is not specified (not known in a
given experiment), the probability should be averaged over
$t_\pi$. This eliminates the oscillation term, but preserves the
{\it global} dependence on neutrino flavour $\nu_a$.
Note, however, that $t_\pi$ was used by us to deduce the position of
the pion decay point $x_i$. If this position is known from other
considerations (e.g. short decay pipe), then $t_{\nu i}$ is fixed,
and the oscillating term is of course retained. \\

When this paper was near its final stage of preparation,
the authors of ref.\cite{SSW} have replaced the original paper
(E.Sassaroli, Y.N.Srivastava and A.Widom {\sl "Charged Lepton
Oscillations"}, hep-ph/9509261; September 1995, 15 pages) by another
one: Y.N.Srivastava, A.Widom and E.Sassaroli {\sl "Lepton Oscillations"},
hep-ph/9509261 v2,
(November 24, 1996; 2 pages) with the same figures and
the same claim formulated in a new abstract: "A simple but general
proof is presented to show that Lorenz covariance and 4-momentum
conservation alone are sufficient to obtain muon oscillations in pion
decay if the recoiling neutrino oscillate."

On December 13 a new paper \cite{WS} by A.Widom and Y.N.Srivastava has
appeared in the {\sl hep-ph} Archive.
In this paper they claim that
muon oscillations associated with mixed neutrino mass matrices
should manifest themselves in the experiment measuring $(g-2)$ --
the anomalous magnetic moment of the muon.
As is known, the $(g-2)$ experiments are not done in coincidence
with detection of neutrinos which accompany production of the muons.
Therefore we believe that no effect associated  with
neutrino mixing  should be seen in $(g-2)$ experiments.

As follows from our discussion, oscillations of correlation
probability may be observed in the two--detector experiments which
measure both charged leptons and neutrinos in
coincidence.\footnote{ One is tempted to apply  arguments developed in this paper not only
to hypothetical neutrino oscillations, but also to the already
observed neutral kaon oscillations and to similar oscillations
of neutral heavy mesons $D^0$, $D^0_s$, $B^0$ and $B^0_s$. The
obvious analogs of the $\pi \to \mu \nu$ decay would be
the decays $\Lambda_c^+ \rightarrow p \bar K^0$,
$\Xi_c^0 \rightarrow \Lambda \bar K^0$, or
 $D_s^+ \to K^+ {\bar K}^0,~D^+ \to {\bar K}^0 \pi^+,~
D^+ \to K^+ {\bar K}^0$, or decays of  $B$--mesons:
 $B^+ \to {\bar D}^0 \pi^+,~
B^+ \to {\bar D}^0 D^+,~B^+ \to {\bar D}^0 D_s^+$.
The corresponding decay widths in the above examples
are larger than the oscillation frequencies,
and therefore the dispersion of momenta is provided by these
decay widths. The description of EPR effect in $\phi$ and
$\Upsilon$ decays in terms of amplitudes has been advocated in
ref.\cite{KS}.

Claims that neutral kaon oscillations must produce oscillations of
$\Lambda$-hyperons, and that the frequency of kaon oscillations
in $\phi$-meson decay is factor of 2 larger than its standard
generally accepted value, have been published in
refs.\cite{SWSPL,SWS330},  and have been disproved in
refs.\cite{LOWE,BK96}.  }
These oscillations would look like an EPR effect, though this
is a simple consequence of standard neutrino oscillations and
relativistic kinematic relations.  Muons do not oscillate.

\newpage

\vspace{2cm}

{\bf Acknowledgments} \\

We are grateful to Mario Greco, who drew attention of one of us
to ref.\cite{SWS}. We thank S.P.Denisov and S.S.Gershtein for the
discussion of IHEP neutrino experiment,
T.Goldman for informing us about some references which were missing
in the preliminary version of this paper,  A.N.Rozanov for the
discussion of CERN neutrino experiment, M.I.Vysotsky for his
role of devil's advocate, and V.L.Telegdi for critical reading of the
manuscript and many helpful suggestions.

This work was partially supported by several grants.

The work by A.D. was supported in part by the Danish National Science
Research Council through grant 11-9640-1 and in part by Danmarks
Grundforskningsfond through its support of the Theoretical
Astrophysical Center.

The work of A.M. is supported by the RFBR Grant 96-15-96939.
He also acknowledges the support of DFG and hospitality of
Humboldt University, Berlin, and IFH, Zeuthen, during the
work on this paper.

L.O. acknowledges the RFBR Grants 96-02-18010 and 96-15-96578 and
Humboldt award.

The work of M.S. is supported by a Grant of the NFR and
INTAS-RFBR Grant 95-605.

\end{document}